\documentclass[prb,twocolumn,showpacs,preprintnumbers,amsmath,amssymb]{revtex4}

\usepackage{graphicx}
\usepackage{epstopdf}
\usepackage{dcolumn}
\usepackage{bm}

\begin{document}


\title{Anomalous scattering in superconducting indium-doped tin telluride}

\author{A. S. Erickson, T. H. Geballe, I. R. Fisher} 
\affiliation{Department of Applied Physics and Geballe Laboratory for Advanced Materials, Stanford University, CA 94305; and Stanford Institute for Materials and Energy Science, SLAC National Accelerator Laboratory, Menlo Park, CA 94025.}
\author{Y. Q. Wu and M. J. Kramer}
\affiliation{Division of Materials Sciences and Engineering, Ames Laboratory and Department of Materials Science and Engineering, Iowa State University, Ames IA 50011.}

\date{\today}

\begin{abstract}
Results of resistivity, Hall effect, magnetoresistance, susceptibility and heat capacity measurements are presented for single crystals of indium-doped tin telluride with compositions Sn$_{.988-x}$In$_x$Te where $0 \leq x \leq 8.4 \%$, along with microstructural analysis based on transmission electron microscopy. For small indium concentrations, $x \leq 0.9 \%$ the material does not superconduct above 0.3 K, and the transport properties are consistent with simple metallic behavior. For $x \geq 2.7 \%$ the material exhibits anomalous low temperature scattering and for $x \geq 6.1 \%$ bulk superconductivity is observed with critical temperatures close to 2 K. Intermediate indium concentrations $2.7\% \leq x \leq 3.8\%$ do not exhibit bulk superconductivity above 0.7 K. Susceptibility data indicate the absence of magnetic impurities, while magnetoresistance data are inconsistent with localization effects, leading to the conclusion that indium-doped SnTe is a candidate charge Kondo system, similar to thallium-doped PbTe. 
\end{abstract}
\maketitle

\section{introduction}

Tin telluride, SnTe, is a small band gap IV-VI semiconductor, and can be doped to degeneracy by either tin or tellurium vacancies.\cite{nimtz} No carrier freeze out is observed to the lowest temperatures, which has been explained for PbTe and closely related materials in terms of vacancy states located deep in the valence/conduction bands respectively.\cite{Parada} For the case of tin deficiency (hole doping), relevant to the current work, the associated Fermi surface comprises light holes located at the $L$-points of the Brillouin zone (using standard nomenclature for the FCC lattice) and, for sufficiently high carrier densities, heavy holes centered near the $\Sigma$ point.\cite{Allgaier} For carrier densities exceeding approximately $10^{20} cm^{-3}$ Sn$_{1-\delta}$Te superconducts, with critical temperatures remaining below 0.3 K,\cite{hulm,hein} which has been interpretted in terms of interband scattering within an otherwise standard BCS mechanism.\cite{allen1969,gubser} However, when doped on the Sn site with In, superconductivity is found as high as 2 K, though the carrier concentration remains similar.\cite{hulm,miyauchi,Bushmarina1986} Notably, indium is known to skip the $+2$ valence, in favor of $+1$ and $+3$, in most materials. For example, the semiconductor InTe comprises two distinct In sites, each with different distances to the ligand Te ions, and is more properly written In(I)In(III)Te$_2$. This tendency leads to speculation as to whether valence fluctuations associated with the In impurities in SnTe might play a role in the enhanced superconductivity. 

Several P-block elements are known to skip valence states in solution or solids. This behavior reflects a combination of the stability of a full/empty S shell, and the formation of chemical bonds with coordinating ligands.\cite{robinday} The effect can be described by a negative effective $U$ within an effective Hamiltonian,\cite{anderson,varma} although the degree to which the attractive interaction reflects relaxation of the coordinating ligands\cite{anderson,walt} versus a more electronic polarization\cite{moizhes1981,moizhes1982} is debated, and depends on the specific materials considered. Several authors have shown that models based on negative-$U$ Hamiltonians yield superconductivity for various ranges of physical parameters,\cite{varma,hirsch,moizhes1987,shelankov1987,schuttler,taraphder,krasinkova1991,malshukov,joerg} indicating that valence skipping elements might provide a credible avenue for the discovery of new superconducting materials. Indeed, reasoning along these lines (rightly or wrongly) has historically lead to the discovery of superconductivity in Ba(Pb$_{1-x}$Bi$_x$)O$_3$ and Ba$_{1-x}$K$_x$BiO$_3$.\cite{sleight,cava} Both of these materials are derived from the charge density wave insulator BaBiO$_3$, which, similar to In(I)In(III)Te$_2$ described above, comprises two distinct Bi sites with different Bi-O bond lengths such that the chemical formula might be more properly considered Ba$_2$Bi(III)Bi(V)O$_6$.\cite{cox} 

Returning to the specific case of In impurities in IV-VI semiconductors, x-ray photoemission spectroscopy measurements of In-doped PbTe have directly revealed the presence of a mixed In valence in that material, comprising both In(I) and In(III) states.\cite{drabkin1982} In-doped PbTe is n-type with a relatively low carrier concentration ($\sim 8 \times 10^{18}$cm$^{-3}$) even though the indium concentration can reach $\sim 20\%$, and transport data have been analysed in terms of pinning of the chemical potential in quasi-local indium impurity states close to the bottom of the conduction band.\cite{ravich2002} The material has an extremely long lived photoconductivity; the barrier to recombination of non-equilibrium carriers presumably reflecting the negative correlation energy of two-electron In traps in this material.\cite{kaidanov1985,volkov2002,drabkin1983} Similar long-lived non-equilibrium effects can be induced by high magnetic fields in alloys (Pb$_{1-x}$Sn$_x$)Te doped with indium.\cite{Brandt} In contrast, the closely related cases of Tl-doped PbTe and In-doped SnTe are both p-type, with considerably higher carrier densities ($\sim 10^{20}-10^{21}$cm$^{-3}$). Tl-doped PbTe does not exhibit non-equilibrium effects,\cite{volkov2002}, and it is reasonable to anticipate that In-doped SnTe would not either. Perhaps significantly, both of these materials do, however, superconduct.

Similar to the case of In-doped SnTe, Tl-doped PbTe is also an anomalous superconductor, with $T_c$ values substantially higher than otherwise anticipated for the range of possible carrier densities.\cite{nemov1998} However, whereas Sn$_{1-\delta}$Te superconducts below 0.3 K, PbTe doesn't superconduct to the lowest temperatures measured when either hole or electron doped by Pb or Te deficiency respectively, or for any other dopant except thallium.\cite{hulm} Hall effect measurements indicate that beyond a threshold concentration, Tl impurities in PbTe are ambivalent, corresponding to a mixture of Tl(I) and Tl(III).\cite{murakami} Significantly, superconductivity is only observed for concentrations beyond this threshold value, for which the Fermi level is effectively pinned.\cite{nemov1998} Various transport measurements have been interpretted in terms of resonant scattering of holes from quasi-localized Tl impurity sites buried beneath the top of the valence band.\cite{kaidanov1985} More recent core level spectroscopy measurements directly confirm the presence of both thallium valences in this regime.\cite{kaminskicomment} Furthermore, the observation of Kondo-like scattering in both resistivity\cite{yana,yanaconfproc} and thermopower\cite{matusiak} measurements, in the absence of magnetic impurities, has been interpretted as evidence for a charge Kondo effect associated with the degenerate Tl(I) and Tl(III) states.\cite{joerg} This behavior can be tuned by additional counterdopants, for which it is still found that the occurance of superconductivity is correlated with both Fermi level pinning and also the observation of anomalous low temperature scattering.\cite{AnnPbTe} Taken together, these observations point towards a picture in which valence fluctuations between degenerate Tl states play a significant role in Tl-doped PbTe, possibly related to the observation of superconductivity in this otherwise non-superconducting material. One would of course like to see whether these ideas are generic to a broader class of material, which is the basic motivation behind studing the closely related case of superconducting In-doped SnTe.  

For a number of reasons, In-doped SnTe is, however, a slightly more complicated material than Tl-doped PbTe. First, SnTe  suffers a structural phase transition,\cite{muldwater1973} the temperature of which depends on the carrier density due to screening of the associated phonon modes.\cite{kobayashi1976,katayama} At one level, this can be seen as an advantage, since the critical temperature of the structural phase transition $T_{SPT}$ can be tuned relative to that of the superconductivity, $T_c$. As a consequence, it can be shown that the anomalously high $T_c$ in this material does not derive from soft phonon modes associated with the structural transition.\cite{myprb} However, since the structural transition affects the temperature-dependence of the resistivity, it is necessary to first completely suppress the structural transition before scattering associated with the In impurities can be studied in isolation. Second, the threshold indium concentration beyond which Fermi level pinning is believed to occur is higher in In-doped SnTe than for Tl in PbTe,\cite{Bushmarina1984} due in part to the large Sn deficiency necessary to suppress the structural transition. Consequently, it is much less clear whether analysis based on a single impurity Kondo model will be appropriate. Third, perhaps related to the larger solid solubility of In in SnTe relative to Tl in PbTe, the behavior of the Hall coefficient for higher indium concentrations is much more complicated in this material. Whereas for Tl-doped PbTe the Hall coefficient rises very slowly beyond a charactristic threshold Tl concentration, leading to a simple picture based on Fermi level pinning, Hall effect data for Sn$_{1-\delta-x}$In$_x$Te have been interpretted in terms of a picture in which the energy of the indium impurities themselves depend on the In concentration, and move deeper in to the valence band for increasing $x$.\cite{Bushmarina1984,Bushmarina1986} The consequent variation in the chemical potential as a function of doping makes a detailed quantitative analysis more challenging.  Regardless of these potential complications, In-doped SnTe provides a clear opportunity to see whether the effects observed in Tl-doped PbTe are generic to this class of impurity in this class of host, and hence to approach the broader question of whether valence skipping elements really provide a credible avenue to the search for new superconductors. For the closely related case of p-type In-doped Pb$_{1-x}$Sn$_x$Te, coincident with pinning of the Fermi level in these indium states, an increase in the scattering rate has been observed via resistivity measurements, \cite{Parfenev,Popov1,Popov2} but to-date the temperature dependence of the resistivity at low temperatures has not been systematically explored.  

In this paper we describe the results of measurements probing the thermodynamic and transport properties of single crystal samples of In-doped SnTe. We have chosen a synthesis technique that deliberately introduces a relatively large Sn vacancy concentration, such that the structural transition is completely suppressed for all indium concentrations studied. We find that, similar to Tl-doped PbTe,  In-doped SnTe also suffers anomalous low-temperature scattering for In concentrations beyond a threshold value. The temperature dependence varies with indium concentration, but for several compositions is certainly reminiscent of the Kondo effect. Susceptibility measurements put an upper limit on the possible concentration of magnetic impurities of $\sim$ 10ppm, indicating that the resistivity upturn does not originate from magnetic Kondo behavior.  Magnetoresistance data show that this effect is largely independent of field up to 14 T, implying that it does not originate from localization effects either. Variation in the carrier density with increasing indium concentration leads to non-monotonic behavior of both the residual resistivity and also the magnitude of the low-temperature resistivity upturn, complicating analysis of the anomalous behavior. Bulk superconductivity is observed above 0.7 K for indium concentrations above 6.1\%, but indium inhomogeneity leads to filamentary superconductivity for lower concentrations. Nanoscale inclusions are observed above 6.5\%, indicating that the solid solubility might be lower than previously thought. Difficulties aside, our results demonstrate that indium impurities in SnTe behave very similarly to Tl impurities in PbTe, and hence that the enhanced superconductivity in both systems is likely related to the same physical mechanism. However, it is also apparent, for the reasons introduced above, that indium-doped SnTe is much less of a model system for quantitative analysis of the putative charge Kondo behavior. 

\section{Experimental methods}
Single crystals of Sn$_{.988-x}$In$_x$Te were grown by an unseeded physical vapor transport method. This method was specifically chosen because it results in a relatively high Sn vacancy concentration, suppressing the structural phase transition so that additional scattering associated with the In impurities can be studied in isolation. For nominal In concentrations, $x_{nom} < 10\%$, starting material in the SnTe:In:Te ratio (1-x$_{nom}$):x$_{nom}$:x$_{nom}$ were ground thoroughly and pressed into a pellet.  The pellet was sintered at 600 $^\circ$C for 36 hrs.  The resulting polycrystalline pellet was ground, pressed, and sintered once more.  For higher In concentrations, a thorough grind before the first sinter step was not possible, as the metal is highly malleable.  In these cases, the starting materials were coarsely ground, pressed, and sintered before it could be thoroughly ground for the second sinter.  For these materials, a third sinter was also performed, to ensure a homogeneous mixture.  For all In concentrations, the resulting pellet was then coarsely crushed, placed at the center of a 10 cm long evacuated quartz tube, and heated to 750 degrees for 9 days under a temperature gradient of 1$^\circ$C/cm.  The resulting single crystals, which grew at the cooler ends of the quartz ampoule, were then annealed 12 hours at 550 C in an inert invironment to improve homogeneity, determined by reproducibility of resistivity measurements. The composition was measured by electron microprobe analysis (EMPA), using PbTe as a tellurium standard, In metal, and Sn metal standards. Material of measured In concentration, $x = 0\%, 0.8\%, 0.9\%, 2.7\%, 3.8\%, 6.1\%, 6.5\%, 7.3\%,$ and $8.4\%$ were analyzed in this way.  Measured concentrations were approximately 80\% of the nominal composition, presumably reflecting loss of indium during the repeated grindings.

Bars were cleaved from single crystals for electrical tranpsort measurements.  Resistivity data were obtained at frequencies of 13.5 Hz, 13.7 Hz or 37 Hz, and current densities along the [100] direction of order 100 mA/cm$^2$ at temperatures above 350 mK and 10 mA/cm$^2$ for data taken below 350 mK.  Data below 1.8 K were taken at a variety of current densities to check for heating effects. Measurements were made for several samples for each composition. Hall measurements were conducted using a Quantum Design Physical Properties Measurement System, at a frequency selected to reduce noise (either 37 Hz, 47 Hz, or 53 Hz) and typical current densities of order 100 mA/cm$^2$.  The Hall coefficient, $R_H$, at a temperature of 5 K was obtained from linear fits to the transverse voltage in fields from -9 to 9 T, aligned along the [001] direction. All electrical contacts were made by sputtering gold contact pads in appropriate geometry, then fixing platinum wire to the pads using Epotek H20E conductive silver epoxy, with typical contact resistances between 2 and 4 $\Omega$. 

Magnetic susceptibility was measured in a Quantum Design MPMS5 SQUID magnetometer, in applied fields of 1000-5000 Oe, depending on sample mass.  Single crystals or groups of single crystals totaling at least 25 mg were held between two plastic straws.   

Heat capacity was measured between 0.35 K and 5 K on 5-12 mg single crystals using the relaxation method with a Quantum Design Physical Properties Measurement System equipped with a helium 3 cryostat.  Single crystals were cleaved to provide a flat surface for good thermal contact to the sample platform.  Measurements were made in zero field and in an applied field of 1 T, to suppress the superconducting transition, at arbitrary orientations of the sample in the field. The Sommerfeld coefficient, $\gamma$, was obtained from linear fits to $C / T$ vs T$^2$ for data taken below 1 K, in an applied field greater than $H_{c2}$ to suppress the superconducting transitions.

Homogeneity was evaluated by transmission electron microscopy.  Samples were glued to a 3mm Cu substrate with a 1 mm hole. They were then polished to $\sim$ 20 $\mu$m thickness and ion milled at $\sim$ -120 $^\circ$C using a liquid nitrogen cooled stage.  Initial ion milling conditions were 5 kV and 5mA at 20 $^\circ$ and then 3 kV and 3 mA at 12 $^\circ$ to perforation.  Evaluation was performed using a FEI Tecnai G$^{2}$ F20 STEM at 200 kV.

\section{Results}

\begin{figure}
\centering
\includegraphics[width=3.5in]{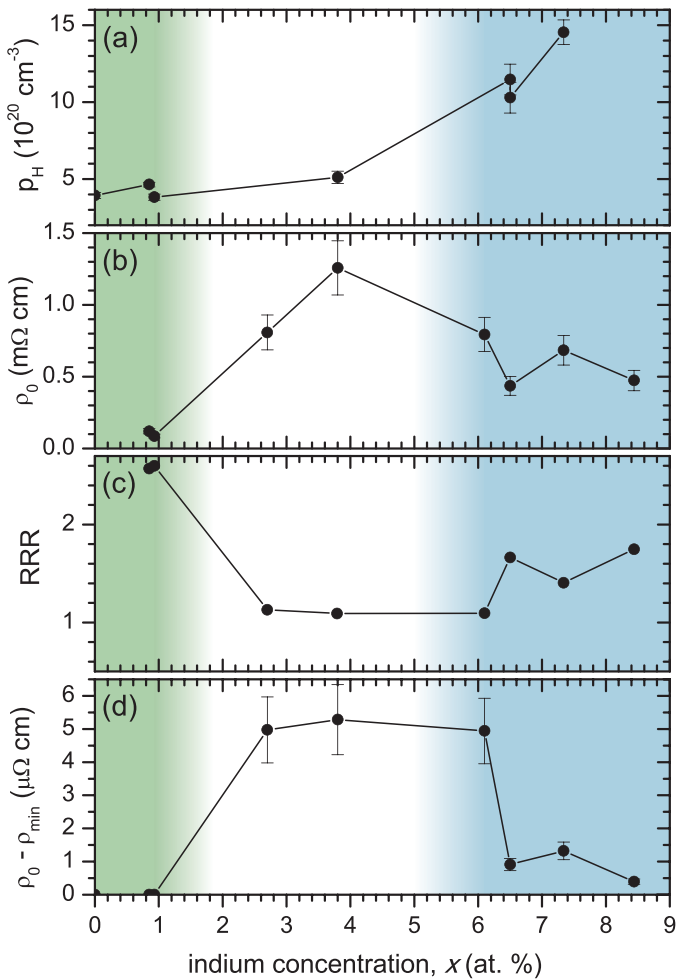}
\caption{\label{resdatavt} (Color online) Summary of data extracted from transport measurements of single crystal samples of Sn$_{.988-x}$In$_x$Te, shown as a function of the measured indium concentration $x$. (a) Hall number, $p_H$ (note that all compositions measured are p-type); (b) residual resistivity $\rho_0$ estimated from the value at 1.8K, the base temperature of the He-4 cryostat used for these measurements; (c) residual resistance ratio, $RRR = \rho(300 K)/\rho_0$;  (d)  $\rho_0 - \rho_{min}$, which provides a measure of the depth of the resistance minimum observed for In concentrations $x \geq$ 2.7. Lines are drawn between data points to guide the eye. The data delineate two broad concentration regimes. For concentrations $x \leq$0.9 at.\% (shaded in green) the material is a good metal, with a relatively low carrier density and no low-temperature upturn in the resistivity. These compositions do not superconduct above 0.3 K. For higher indium concentrations the residual resistivity increases, and a low temperature resistivity upturn is observed. Both $\rho_0$ and the magnitude of the resistivity upturn do not vary monotonically with the indium concentration, but decrease for $x \geq 6.1\%$. Perhaps significantly, bulk superconductivity is only observed above 0.7 K for this higher concentration regime (shaded in blue, and described in greater detail in figure \ref{cpdatavt}).}
\end{figure}

The Hall coefficient $R_H$ of indium-doped tin telluride is positive, indicating p-type carriers. Although tin telluride is a multiband semiconductor, the Hall number, $p_H = 1/R_He$, shown in figure \ref{resdatavt}(a) as a function of indium content, $x$, provides a reasonable estimate of the carrier concentration.\cite{Bushmarina1984}  The tin vacancy concentration, $\delta$, can be estimated from the Hall number of indium-free single crystals of Sn$_{1-\delta}$Te, grown following the same method as the indium-containing samples, assuming each tin vacancy introduces two holes.\cite{Bushmarina1984}  Tin telluride grown in this way has a Hall number, $p_H = 3.9(2) \times 10^{20}$ cm$^{-3}$, indicating a tin vacancy fraction of $\delta = 1.2\%$.  The following analysis assumes a constant value of $\delta$ across the series, but the broad conclusions are not dependent upon this. 

The Hall data shown in figure \ref{resdatavt}(a) are in broad agreement with extensive previous measurements of polycrystalline samples in which both the Sn vacancy concentration and the indium concentration were systematically varied.\cite{Bushmarina1984} This earlier study concluded that the Fermi level is pinned in resonant indium impurity states (below the top of the valence band) for vacancy concentrations $\delta \leq x/2$.\cite{Bushmarina1984} For the estimated $\delta = 1.2\%$ of our samples, we can therefore anticipate that the Fermi level is pinned in the indium levels for $x \geq 2.4$ \%. Variation in the Hall coefficient as a function of indium concentration for a fixed Sn vacancy concentration was accounted for by the authors of ref.\onlinecite{Bushmarina1984} in terms of a dependence of the energy of the indium impurity levels with the indium concentration, such that the indium levels move deeper in to the valence band with increasing $x$. The increased value of $p_H$ shown in figure \ref{resdatavt}(a) for higher values of $x$ are consistent with this previous analysis. 

\begin{figure}
\centering
\includegraphics[width=3.5in]{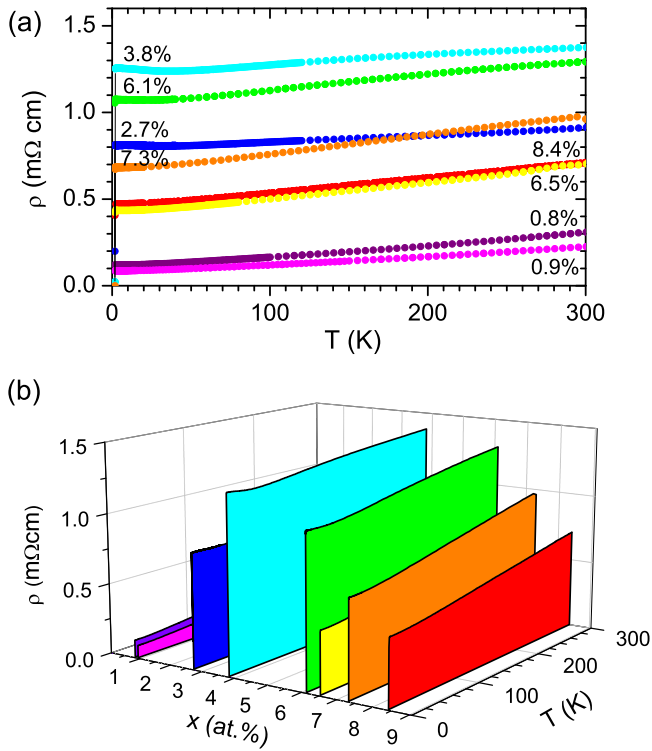}
\caption{\label{rhosnteinvt} (Color online) Temperature dependence of the resistivity of single-crystal samples of Sn$_{.988-x}$In$_x$Te in zero applied field. (a) Individual curves are shown as a function of temperature and labeled by the indium concentration, $x$. (b) The same data are plotted as a function of both temperature and the indium concentration to more clearly reveal the non-monotonic doping-dependence of the indium impurities. The same color scheme is used in both panels. The data establish a general trend in which the residual resistivity first rises with increasing indium concentration before falling again as $x$ is increased beyond $3.8\%$. Data for $x = 6.5\%$ (in yellow) appear to be anomalously low, most probably associated with errors in the geometric factors used to calculate absolute values of the resistivity.}
\end{figure}

\begin{figure}
\centering
\includegraphics[width=3.5in]{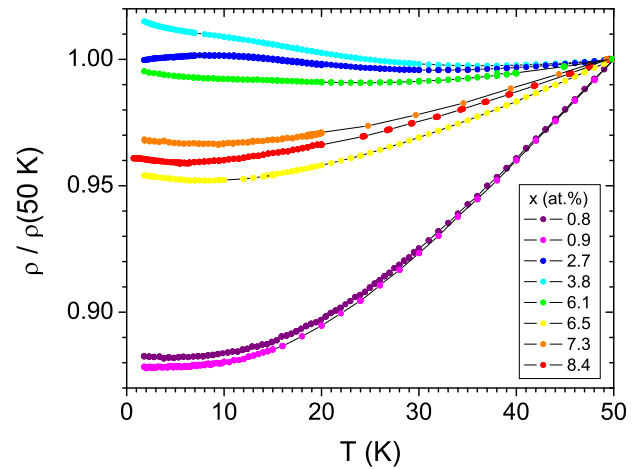}
\caption{\label{normres} (Color online) Temperature dependence of the resistivity between 0 and 50 K, illustrating the low-temperature upturns observed for indium concentrations $x \geq$ 2.7 at.\%. Data are normalized by their value at 50 K so that all curves can be fit on to a single graph. The same color scheme is used as in Figure 2. Absolute values of the residual resistivity and the magnitude of the resistivity upturn can be read from Fig 1(b) and (d) respectively. Data were collected in an applied field of 1 T to suppress the superconducting transition.}
\end{figure}

Representative data showing the temperature dependence of the resistivity between 1.8 K and 300 K for Sn$_{.998-x}$In$_x$Te, with $0.8\% \leq x \leq 8.4\%$, are shown in figure \ref{rhosnteinvt}. The data clearly reveal a non-monotonic dependence on the indium concentration, most readily appreciated by inspection of panel (b). For In concentration $x \leq 0.9\% $ the resistivity decreases monotonically with decreasing temperature, and the residual resistivity ($\rho_0$, shown in figure \ref{resdatavt}(b), estimated from the value of the resistivity at 1.8K since the T-dependence is very weak at low temperatures) is approximately 100$\mu\Omega$cm. The residual resistance ratio, $RRR = \rho(300 K)/\rho_0$ for these low indium compositions is 2.6 (figure \ref{resdatavt}(c)). The range of compositions for which this general good-metallic behavior is observed are shaded green in figure \ref{resdatavt}. There is no signature of a phase transition in the resistivity or its derivatives) for these compositions, consistent with complete suppression of the structural phase transition. Higher indium concentrations, corresponding to $x \geq 2.7\%$, reveal a much larger residual resistivity (as large as 1.3 m$\Omega$cm for $x$=3.8\%), and correspondingly smaller values of the $RRR$. Given that the Hall coefficient for the $x$=0.9 and 3.8\% samples are almost identical, the large increase in the resistivity clearly reflects a dramatic increase in the scattering rate, which has previously been attributed to resonant scattering from indium impurity levels.\cite{Bushmarina1984} The residual resistivity does not, however, increase monotonically with increasing indium concentration, but decreases for $x > 6.1\%$. Comparing panels (a) and (b) of figure \ref{resdatavt}, the factor by which the residual resistivity decreases as $x$ is increased beyond 3.8\% is almost identical to that by which the Hall number $p_H$ increases, at least within the uncertainty of the measurements, which are dictated primarily by the geometric factors, and hence up to approximately 20\%. Hence, if to a first approximation we assume that the effective mass is independent of $x$, it appears that the primary cause for the reduction of the residual resistivity as $x$ increases beyond 3.8\% is the larger carrier concentration, which has been attributed to the progressive displacement of the indium impurity levels deeper in to the valence band,\cite{Bushmarina1984} and not from a change in the scattering rate. 

\begin{figure}
\centering
\includegraphics[width=3.5in]{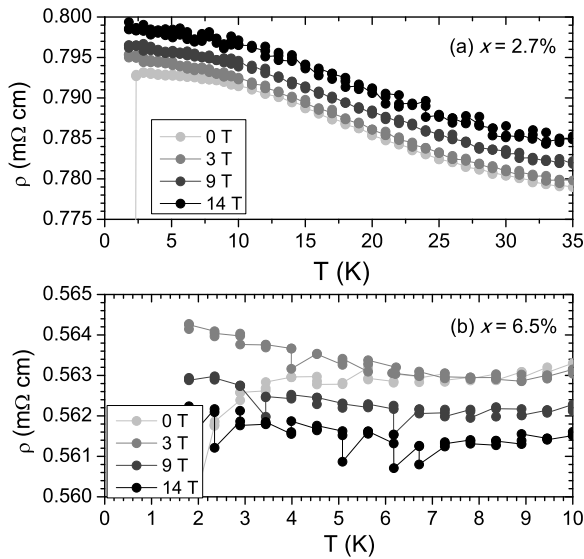}
\caption{\label{magresvst} Temperature dependence of the resistivity for (a) $x=2.7\%$ and (b)$x=6.5\%$ for transverse fields of 0, 3, 9 and 14 T. Notably, after superconducting fluctuations have been suppressed, the temperature dependence of the resistivity upturn is apparently unaffected by fields up to 14 T.}
\end{figure}

Inspection of the data shown in figure \ref{rhosnteinvt} reveal that for $x \geq 2.7\%$, the resistivity turns up at low temperatures. This effect is more clearly seen in Figure \ref{normres}, showing data for the same samples in an applied field of 1 T to suppress the superconducting transitions, and normalized at 50 K to emphasize the low-T behavior. For $x = 2.7\%$, a low temperature resistance minimum is observed at 30 K, followed by a broad maximum near 10 K. The subsequent downturn varies between samples (see for example the zero-field data in figure \ref{magresvst}(a)), and may be related to superconducting fluctuations of an impurity phase, discussed later.  For $x = 3.8\%$, a deeper minimum at a temperature of 40 K is present, followed by a shoulder at 10 K, and a further increase in resistivity as the temperature decreases further.  For $x = 6.1\%$,  the temperature at which the resistance minimum occurs is closer to 20 K.  For $6.5\% \leq x \leq 8.4\%$, the magnitude of the resistivity upturn decreases, with a minimum centered at around 8 K. A crude estimate of the magnitude of the resistivity upturn is provided by the difference of the residual resistivity and the minimum value of the resistivity, $\rho_0 - \rho_{min}$, shown in panel (d) of figure \ref{resdatavt} as a function of indium concentration.  

\begin{figure}
\centering
\includegraphics[width=3.5in]{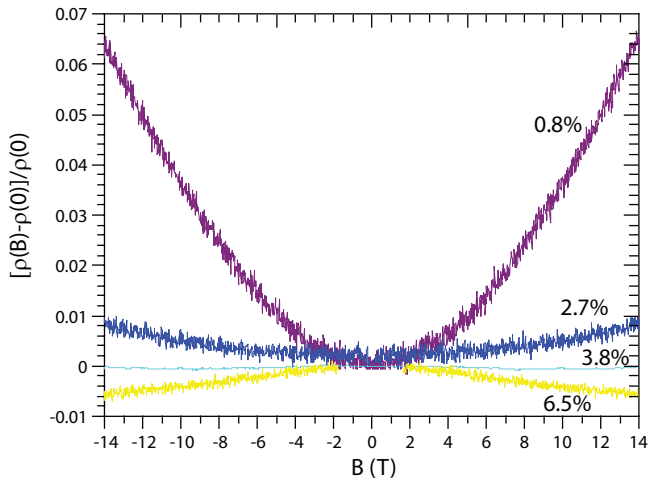}
\caption{\label{mrsnteinvt} (Color online) Field dependence of the transverse magnetoresistance, $[\rho(B) - \rho(B = 0)]/\rho(B = 0)$, at 0.5 K for Sn$_{.988-x}$In$_x$Te with $x = 0.8\%, 2.7\%, 3.8\%$, and $6.5\%$. For $x$=6.5\%, data are only shown for fields greater than $H_{c2} \sim 1T$. Data for lower fields are shown in figure \ref{hc2}. The same color scheme is used as in figures 2 and 3.}
\end{figure}

The origin of the resistivity upturn was probed via magnetoresistance measurements. In all cases, the field was applied perpendicular to the current. Data were taken both as a function of field and temperature. Representative data showing field sweeps at a fixed temperature of 0.5 K are shown in figure \ref{mrsnteinvt} for four In concentrations. A small linear contribution to the magnetoresistance, odd with respect to the applied field (hence originating from mixing of the Hall signal), was present for most samples, and has been subtracted to generate the plots shown in figure \ref{mrsnteinvt}. Low concentration samples, with $x = 0.9\%$, show typical metallic behavior, with the resistivity increasing with the square of the applied field at low fields.  Material with $x = 2.7\%$ show positive, but shallow magnetoresistance.  For $x = 3.8\%$, the magnetoresistance is nearly flat.  For $x = 6.5\%$, bulk superconductivity is observed at low fields, and is fully suppressed by $H = 1$T (discussed in greater detail below).  For fields above this value, negative magnetoresistance is observed, with a linear field dependence for the field range measured.

Measurements of the temperature dependence of the resistivity in fixed magnetic fields provide further insight. Data for two specific indium concentrations ($x$ = 2.7 and 6.5 \%, corresponding to compositions which show positive and negative magnetoresistance in figure \ref{mrsnteinvt} respectively) are shown in figure \ref{magresvst} for applied fields of 0, 3, 9 and 14 T. These data reveal that the field dependence of the magnetoresistance shown in figure \ref{mrsnteinvt} is not due to suppression or enhancement of the resistivity upturn. Rather, it appears that the magnitude of the upturn for the two compositions shown remains essentially unchanged for the field ranges studied once superconducting fluctuations have been suppressed by fields of 3 T. The origin of the negative magnetoresistance for the highest indium concentration is unclear, especially in light of the fact that these compositions have a lower residual resistivity than the intermediate indium concentrations for which the magnetoresistance is positive.

\begin{figure}
\centering
\includegraphics[width=3.5in]{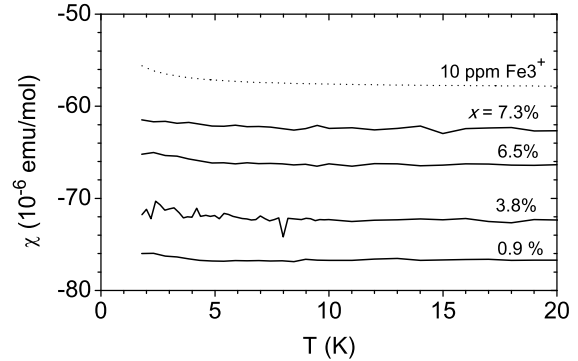}
\caption{\label{chidatavt} Temperature-dependence of the magnetic susceptibility at low temperature for four representative indium concentrations (mol refers to the formula unit). Measurements were made in applied fields of 1000 to 5000 Oe. The calculated Curie susceptibility which would arise from 10 ppm Fe$^{3+}$ impurities is drawn as a dotted line, having subtracted a T-independent offset for easy comparison with the diamagnetic susceptibility of the In-doped SnTe samples.}
\end{figure}

The normal state behavior was also investigated by measurements of the magnetic susceptibility. Results are shown in figure \ref{chidatavt}.  The material is diamagnetic, consistent both with the overall low density of states and with expectations that indium impurities adopt a formal valence corresponding to the closed-shell ion configurations of either $+1$ or $+3$.  The material becomes less diamagnetic with increasing indium concentration, consisent with the increased density of states inferred from heat capacity measurements (described below).  Significantly, within the experimental resolution, no systematic Curie-like response is evident in the data above 1.8 K. For comparison, the expected behavior of 10 ppm Fe$^{3+}$ impurities is also shown.  This is above, but near, the resolution limit of the instrument at these temperatures, indicating that the presence of magnetic impurities can therefore be limited to less than 10 ppm Fe$^{3+}$.

\begin{figure}
\centering
\includegraphics[width=3.5in]{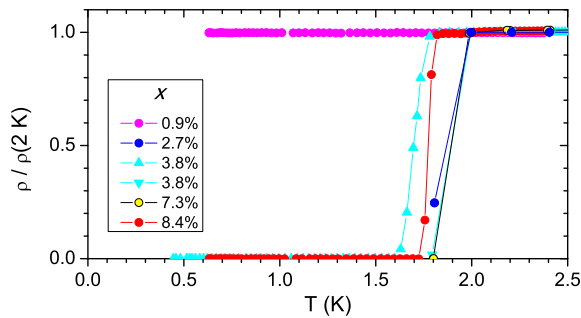}
\caption{\label{restrans} (Color online) Superconducting transitions observed in the resistivity for samples with indium concentration $x \geq 2.7\%$. Transitions for samples with $x$ = 2.7\% and 7.3\%, and one sample with $x$ = 3.8\% (downward pointing triangles), lie almost on top of each other, partially obscuring the data points. The same color scheme is used as in figures 2 and 3.}
\end{figure}

\begin{figure}
\centering
\includegraphics[width=3.8in]{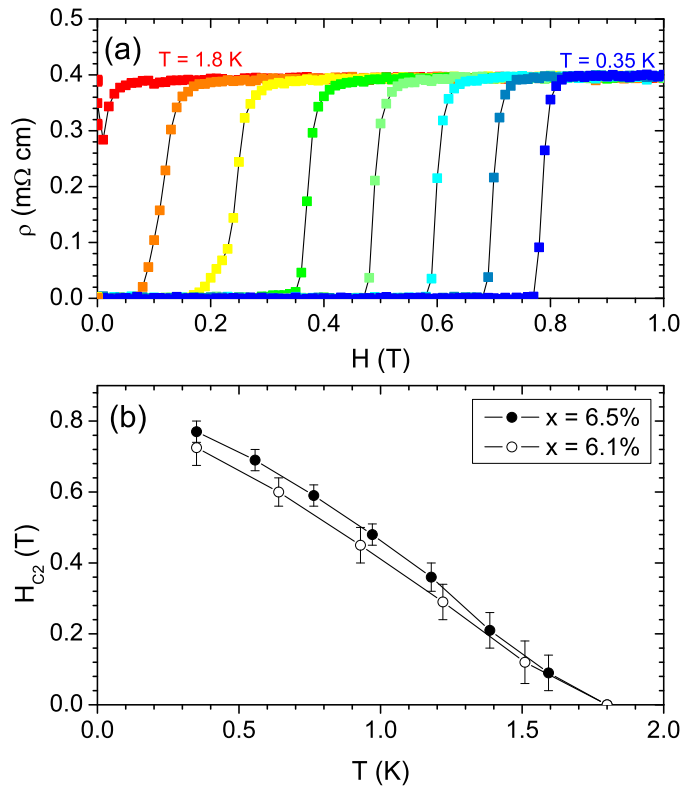}
\caption{\label{hc2} (Color online) (a) Field dependence of the resistive transition for $x$ = 6.5\% for temperatures between 0.35 K (blue) and $T_c$ $\sim$ 1.8 K (red). (b) $H_{c2}$ extracted from resistive transitions for $x$ = 6.5\% (solid symbols, extracted from the data shown in panel (a)) and $x$ = 6.1\% (open symbols) as described in the main text. Lines are drawn between data points to guide the eye.}
\end{figure}

\begin{figure}
\centering
\includegraphics[width=3.7in]{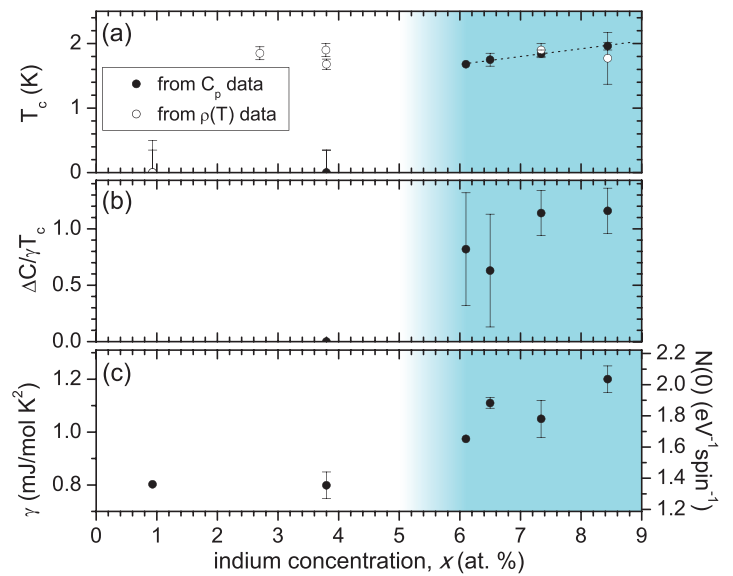}
\caption{\label{cpdatavt} (Color online) Summary of data extracted from heat capacity measurements of Sn$_{.988-x}$In$_x$Te. (a) Superconducting critical temperatures derived from heat capacity measurements (closed circles) and resistive transitions (open circles) as a function of In concentration, $x$.  Dashed line between data is drawn as a guide to the eye. (b) Magnitude of the normalized superconducting anomaly, $\Delta C/\gamma T_c$, compared to the value predicted by BCS theory for weakly coupled superconductors, 1.43, represented by the top axis. (c) Sommerfeld coefficient, $\gamma$ (left axis), and corresponding density of states at the Fermi level, $N(0)$ (right axis). The data delineate two concentration regimes. For indium concentrations $x \geq$ 6.1 at. \%, bulk superconductivity is observed above 0.7 K (blue shaded region). For $x \leq$ 3.8 at. \%, no bulk superconductivity above 0.7 K is observed.}
\end{figure}

All samples with $x > 2.7\%$ measured to low temperature show a transition to zero resistance between 1.5 and 2.0 K (figure \ref{restrans}). Critical temperatures, $T_c$, defined as the midpoint of the resistive transitions, are shown as a function of $x$ in figure \ref{cpdatavt}(a) as open circles. Error bars are defined as 10\% and 90\% of the full transition height. For $x \geq 6.1\%$ the superconducting transition is smoothly suppressed by application of an external field (figure \ref{hc2}), leading to estimates of $H_c2(0) \sim 1T$, in keeping with previous reports.\cite{Parfenev} In contrast, as can be seen from figure \ref{mrsnteinvt}, the superconductivity is destroyed for almost arbitrarily small fields for intermediate indium concentrations $x$ = 2.7 and 3.8 \%, indicating that the resistive transitions observed for these compositions might be associated with a minority second phase. In such a case, filamentary paths presumably percolate through the host matrix due to Josephson coupling between superconducting regions. Phase coherence through these Josephson junctions is destroyed by small magnetic fields, such that the zero-resistance state is suppressed by fields considerably smaller than $H_{c2}$ of the impurity phase. To investigate whether the resistive transitions are associated with bulk superconductivity, additional heat capacity measurements were also performed.

\begin{figure}
\centering
\includegraphics[width=3.8in]{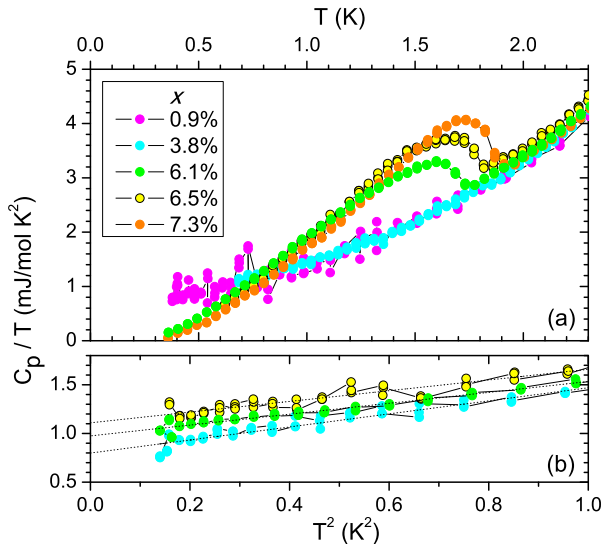}
\caption{\label{cpsnteinvt} (Color online) Representative low temperature heat capacity data of Sn$_{.988-x}$In$_x$Te for several indium concentrations. (a) $C_p/T$ vs. $T$, in zero field, showing the bulk superconducting transition for samples with indium concentration $x \geq$ 6.1 at. \%. (b) $C_p/T$ vs. $T^2$ in an applied field of 1 Tesla, to suppress the superconducting transition. Dashed lines show linear fits to the data below 1K, revealing the Sommerfeld coefficient, $\gamma$. The same color scheme is used as in Figures 2 and 3.}
\end{figure}

Heat capacity data for several representative indium concentrations are shown in figure \ref{cpsnteinvt}(a) as a function of temperature in zero applied field.  $T_c$ values were extracted from the midpoint of the transition, with error bars defined as $90\%$ to $10\%$ of the full transition height (estimated from standard geometric constructions).  The dependence of $T_c$ on In content, $x$, is shown in figure \ref{cpdatavt}(a) as solid symbols.  Significantly, material with $x \leq 3.8\%$ do not show superconducting anomalies in heat capacity data down to 0.7 K (the lowest temperature measured for this composition).  These data confirm that the resistive transitions observed for samples with $x =$ 2.7 and 3.8\% between 1.6 and 1.9 K (open symbols in figure \ref{cpdatavt}(a)) result from trace amounts of superconductivity in an inhomogeneous material.  In contrast, samples with $x \geq 6.1\%$ show superconducting anomalies with a normalized magnitude, $\Delta C/\gamma T_c$, near the BCS prediction for weakly-coupled superconductors (figure \ref{cpdatavt}(b)), demonstrating that the bulk of the material is superconducting for these In concentrations.  

The Sommerfeld coefficient was obtained from linear fits to $C/T$ vs $T^2$ below a temperature of 1 K, shown in figure \ref{cpsnteinvt}(b). Estimates are shown as a function of $x$ in figure \ref{cpdatavt}(c). The corresponding density of states at the Fermi level, $N(0)$, shown on the right axis, is similar for material with $x = 0.9\%$ and $x = 3.8\%$, and begins to increase for $x > 6.1\%$.  Values are similar to those of In-free Sn$_{1-\delta}$Te with comparable carrier concentrations reported in the literature.\cite{phillips1971}

\begin{figure}
\centering
\includegraphics[width=2.5in]{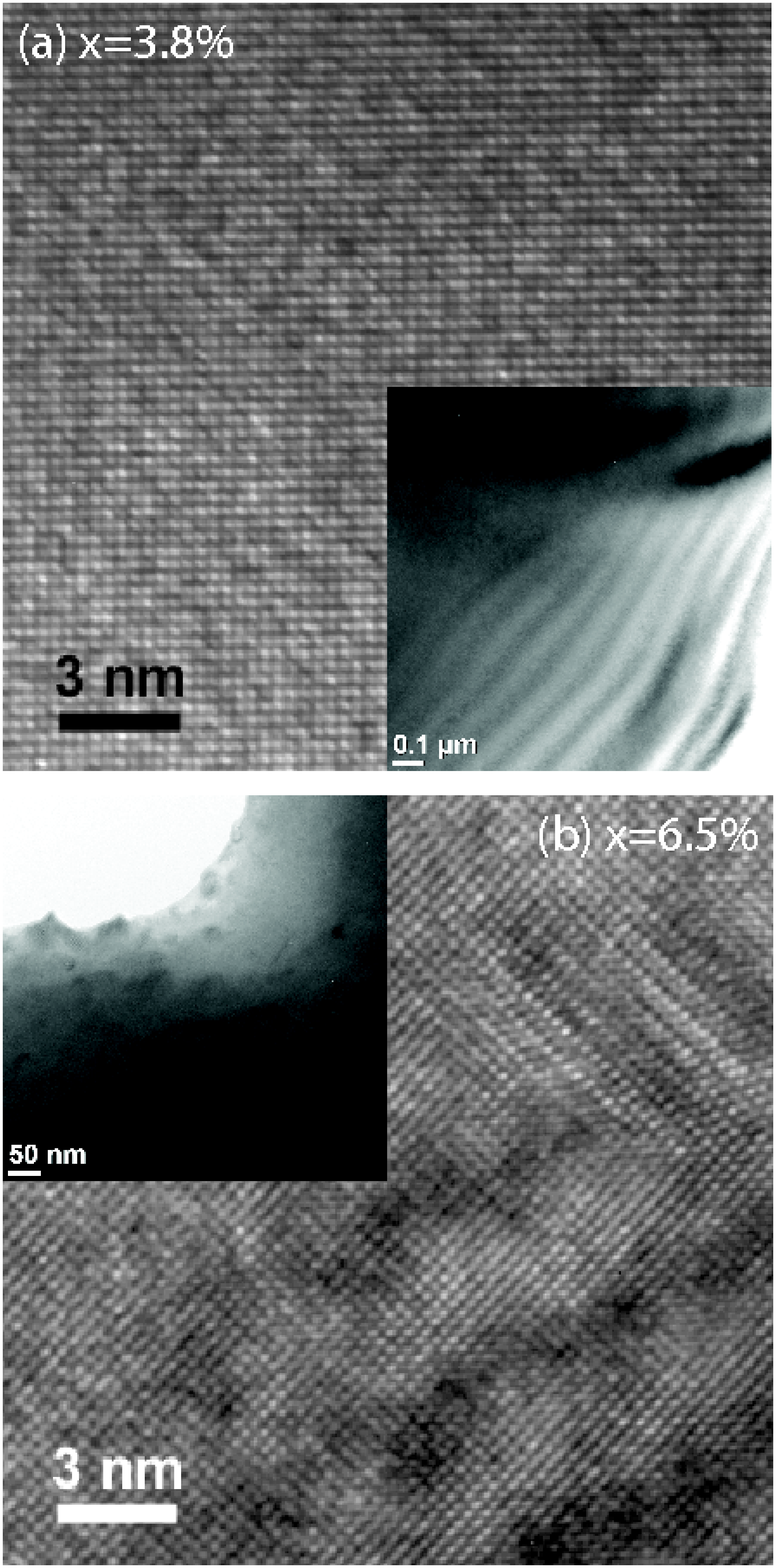}
\caption{\label{tem} High resolution TEM micrographs for single crystals of Sn$_{.988-x}$In$_x$Te along the [001] zone axis for (a) $x = 3.8\%$ and (b) $x = 6.5\%$. In both cases, insets show a lower magnification bright-field image. For $x$ = 3.8\%, the lattice is very well ordered over large length scales. The wider field of view reveals no inclusions. Broad fringes with a 100 nm period correspond to progressive changes in the thickness of the specimen. For the higher indium concentration $x$ = 6.5\% the low magnification image clearly reveals the presence of inclusions with linear dimensions approximately 10 - 20 nm, visible as dark contrast in the bulk matrix. The high resolution TEM image, taken over a region containing an inclusion, illustrates the semi-coherent nature of the defects, resulting in Moire fringes in the lower right and upper right part of the image with approximately a 5 nm period.}
\end{figure}

The presence of superconducting transitions in resistivity measurements, but no bulk transition in specific heat measurements, in samples with $x = 2.7\%$ and $3.8\%$, indicates the presence of inhomogeneity of one form or another. With this in mind, samples were characterized by transmission electron microscopy. Representative data for samples with $x = 3.8\%$ and $x = 6.5\%$ are shown in figure \ref{tem}.  High resolution images for the sample with $x = 3.8\%$ (panel (a) of figure \ref{tem}) reveal a very well-ordered lattice, with no visible defects over extended length scales. Lower magnification bright field images (for example the inset to figure \ref{tem}) show no indication of any inclusions. The absence of any obvious second phase indicates that the minority phase inducing superconducting transitions in resistivity measurements for these intermediate indium concentrations arise from variations in the indium concentration in the host matrix. Such a variation was not apparent in microprobe measurements, indicating either that the variation is on a length scale below 1 $\mu$m, or, more likely, that the relative volume associated with these regions was so small as to be easily missed (associated perhaps with the edges of crystals). In contrast, data for the higher indium concentration, $x = 6.5\%$ , for which bulk superconductivity is inferred from heat capacity measurements, reveal nano-scale precipitates (inset to panel (b)). These precipitates are semi-coherent with the bulk matrix, such that high resolution images of the precipitates reveal Moire fringes due to the two almost identical lattice periodicities (figure \ref{tem}(b)). It is not clear what phase is responsible for these inclusions -- one possibility is cubic InTe, ordinarily stable only at high pressure, perhaps stabilized here by the host matrix. However, the presence of these inclusions certainly signals proximity to the solid-solubility limit of indium in the tin telluride host. The high resolution TEM image is a projection along the [001] zone axis, and as such the precipitates observed in figure \ref{tem}(b) appear to be nearly coherent with the lattice in this orientation. The observation of Moire fringes directly establishes that the precipitates do not extend continuously across the thickness of the thinned TEM specimen, since this effect requires two closely matched periodic lattices to be stacked vertically above one another. The current data do not establish whether the precipitates are thin lamellae or more 3-dimensional inclusions, and as such do not permit an accurate estimate of the volume fraction. However, it is clear that the volume fraction is very small in comparison to the host matrix, indicating that the bulk superconductivity inferred from heat capacity measurements (figure \ref{cpsnteinvt}) is due to the host indium-doped SnTe and not to the precipitates themselves. This is confirmed by the high value of $H_{c2}$ (figure \ref{hc2}) for these higher indium concentrations, since Josephson coupling between non-percolating superconducting precipitates would be suppressed for considerably lower fields. The observation of such a high $H_{c2}$ unambiguously demonstrates that the superconductivity arises from the bulk matrix.

\section{Discussion}

Variation of the Hall coefficient of polycrystalline Sn$_{1-\delta-x}$In$_x$Te has previously been attributed to the competing effects of Sn vacancies ($\delta$, which add two holes per vacancy) and the ambivalent nature of In impurity levels, which have been argued to lie within the valence band.\cite{Bushmarina1984,Bushmarina1986} When $x < 2\delta$ the In impurity concentration is too low to compensate the Sn vacancies, and the Fermi energy lies below the In impurity levels in the valence band. For $x = 2\delta$ the indium concentration is large enough to compensate the Sn vacancies. Larger indium concentrations result in pinning of the Fermi energy in the impurity levels due to the ambivalent nature of indium, as has been observed in other similar systems containing a sufficient concentration of this class of impurity, including In-doped PbTe \cite{drabkin1982,volkov2002} and Tl-doped PbTe \cite{kaidanov1985,nemov1998}. Whereas the Tl impurity levels in PbTe do not appear to drift much with changing impurity concentrations (perhaps because the solubility limit is so low), they do appear to move deeper in to the valence band with increasing $x$ in indium-doped SnTe, such that the Hall number increases by a relatively large amount for larger values of $x$ (figure \ref{resdatavt}(a)).  Analysis of the Hall coefficient of our vapor-grown Sn$_{1-\delta}$Te crystals indicates that $\delta$ = 1.2\% and hence that the Fermi level should lie in the In impurity levels for $x > 2.4\%$. The dramatic increase in residual resistivity for a similar hole concentration that occurs on going from $x = 0.9\%$ to $x = 2.7\%$ is certainly consistent with such a scenario, and indeed indicates that $\delta$ does not vary substantially with $x$, at least up to this concentration. 

Significantly, all compositions with $x \geq 2.7\%$ exhibit not only an enhanced scattering rate but also a low temperature resistivity upturn. The functional form of the upturn varies somewhat with $x$, perhaps related to differences in the temperature-dependence of the normal metallic resistivity, or perhaps reflecting genuine differences in the low-temperature scattering. Several of the data sets are very suggestive of the Kondo effect, (see for example figure\ref{magresvst}(a)). However, the data do not permit a fit, due in part to the poorly characterized T-dependence of the background, and in part to the relatively short temperature interval over which the upturn is observed. Nor can the unitary scattering limit be analyzed with any degree of confidence due to uncertainty in basic materials parameters, including the effective mass and the band offset, in this multiband material. The susceptibility data shown in figure \ref{chidatavt} imply a concentration of magnetic impurities below approximately 10 ppm, and it is therefore highly unlikely that this behavior arises from a magnetic Kondo effect associated either with unpaired spins on the indium impurities (i.e. In$^{2+}$) or with extraneous magnetic impurities. Furthermore, magnetoresistance data do not scale with the residual resistivity (figure \ref{mrsnteinvt}) and in fact the upturns appear to be essentially insensitive to to applied fields up to 14 T (figure \ref{magresvst}), making analysis based on localization equally unlikely. 

Given previous suggestions of a charge Kondo effect in Tl-doped PbTe,\cite{yana} it is of course interesting to consider the possibility that In-doped SnTe hosts a similar effect. Within the charge Kondo model developed by Dzero and Schmalian,\cite{joerg} Tl impurities are described by a Hamiltonian with a negative effective $U$ and are coupled to the bath of conduction electrons with some energy scale $V$. In the limit $U > V$, the problem can be projected on to states corresponding to the presence/absence of pairs of electrons on the impurity sites. This in turn can be described by a pseudospin 1/2 model, in which coherent tunneling of pairs of electrons on/off the impurity sites is described by a pseudospin-flip.\cite{taraphder} The effective Hamiltonian derived from this treatment supports a charge analog of the Kondo effect, in this case describing screening of the pseudospin for temperatures below a characteristic temperature $T_K$.\cite{joerg} Within such a picture, the magnitude of the resistivity upturn arising from Kondo scattering $\rho_{imp}$ for $T < T_K$ can, at least in principal, be analysed in terms of standard unitary scattering theory, leading to an estimate of the number of Kondo impurities $c_{imp}$ via the relation:
\begin{equation}
\rho_{\mathrm{imp}}(0) = 2 m c_{\mathrm{imp}} / [ne^2 \pi \hbar g(E_F)]
\end{equation}
where $m$ is the effective mass, $n$ the hole concentration, and $g(E_F)$ the
density of states at the Fermi level. Initial calculations for Tl-doped PbTe using known band parameters for undoped PbTe lead to estimates well below the actual Tl concentration, possibly implying that not all Tl impurities are able to fluctuate.\cite{yana} This might, for instance, arise if the Tl impurity levels were distributed over a range of energies, such that the two valence states are only degenerate for a limited subset of impurities for any given value of the Fermi energy. However, several factors contribute a relatively large uncertainty to this estimate. In particular, it is not clear whether the band offset between the light and heavy hole pockets in PbTe varies with Tl concentration. In the case of In-doped SnTe, greater uncertainty in materials parameters, including estimates of the effective mass, make such an analysis even more difficult. However, we can at least comment on the non-monotonic dependence of the magnitude of the resistivity upturn, crudely parametrized by the quantity ($\rho_0 - \rho_{min}$) in figure \ref{resdatavt}(d), as the indium concentration is increased beyond 6.1\%. Specifically, as can be appreciated from equation 1, the combined effects of the increased carrier density (figure \ref{resdatavt}(a)) and the increased density of states (figure \ref{cpdatavt}(c)) both act to reduce the $\rho_{imp}$, all else being equal. If the concentration of Kondo impurities is less than the actual indium concentration, as has been suggested for Tl-doped PbTe, and also varies relatively slowly with doping, then it is possible that this effect outweighs any variation in $c_{imp}$ or $m$. Indeed, the relative reduction in ($\rho_0 - \rho_{min}$) on going from $x$=3.8\% to 6.5\% is very similar to the estimated reduction in $\rho_{imp}$ if $c_{imp}$ and  $m$ are unchanged. The relatively slow variation in $T_c$ with $x$ for $x \geq 6.1\%$ is perhaps also indicative of a slow variation in $c_{imp}$ with $x$ within a Charge Kondo picture. This analysis is not intended to be convincing at a quantitative level, but is illustrative that the complicated nonmonotonic variation in the magnitude of the resistivity upturn with increasing indium concentration might be understood within a Kondo picture given the variation in the Hall coefficient and density of states that is known to occur.  

Although it is tempting to consider In-doped SnTe as a candidate charge Kondo system, analysis of the heat capacity data
shown in figure \ref{cpdatavt} indicate that the Sommerfeld coefficient is approximately equivalent to that of Sn$_{1-\delta}$Te with similar carrier densities.\cite{phillips1971} The absence of an obvious enhancement might reflect changes in the band parameters induced by the relatively large concentration of indium impurities, and quantitative analysis is further complicated without an estimate of either the Kondo temperature or the concentration of Kondo impurities. Equally, the functional form of the resistivity upturn for some concentrations (for example that of $x$ = 3.8\% in figure \ref{normres}) appear rather different to that of the traditional single-ion Kondo effect. Although this difference might reflect the temperature dependence of the background metallic contribution, it is equally possible that the relatively large indium concentrations required to obtain pinning of the Fermi level in In-doped SnTe lead to deviations from simple single-ion behavior. In this case, interactions leading to either charge ordering (analogous to magnetic order in the magnetic Kondo effect) or even coherence effects,\cite{taraphder} may play a role. 

Within the charge Kondo picture developed by Dzero and Schmalian,\cite{joerg} and also following similar work by other authors,\cite{varma,hirsch,moizhes1987,shelankov1987,schuttler,taraphder,krasinkova1991,malshukov} valence fluctuations described within an effective negative-$U$ model that lead to Kondo-like scattering can also lead to superconductivity. Describing the two valence states as the 'up' and 'down' states of a pseudo-spin $\frac{1}{2}$, the superconducting state corresponds to XY-order (analogous to a canted XY antiferromagnetic state). Although such ordering competes with the charge Kondo state (which acts to screen the pseudo-spin), the XY order is unfrustrated, and hence is not affected by variation in the distance between dopant ions. For Tl-doped PbTe, the onset of Kondo-like features in the resistivity as a function of composition is indeed observed to correlate with the onset of superconductivity. Previous resistivity measurements of indium-doped SnTe appeared to show a similar correlation between enhanced scattering and superconductivity. \cite{Bushmarina1984,Bushmarina1986} However, as we have shown, bulk superconductivity is only observed above 0.7 K for indium doped SnTe for $x \geq 6.1\%$, implying that resistive transitions observed for intermediate indium concentrations are related to inhomogeneity. Of course we cannot rule out the occurence of bulk superconductivity below 0.7 K for $x$ = 3.8\%. The relatively rapid reduction in carrier density (figure \ref{resdatavt}(a)) and density of states (figure \ref{cpdatavt}(c)) as $x$ is reduced below $\sim$ 6.1\% implies that the modest variation in $T_c$ for $x \geq 6.1\%$ (figure \ref{cpdatavt}(a))might become somewhat sharper for lower concentrations. Even so, it remains to be firmly established whether a correlation between anomalous scattering and superconductivity extend to intermediate indium concentrations $2.7 \leq x \leq 3.8 \%$.

The above data establish In-doped SnTe as another candidate charge Kondo system alongside its nominally isoelectronic counterpart Tl-doped PbTe. The anomalous scattering observed in these two materials appears then to be a generic effect associated with this class of impurity (group III valence skipping elements) in this class of host (degenerate IV-VI semiconductor). There are however several differences between the two materials, some of which make In-doped SnTe a more attractive material to study, others of which make it less so. On the positive side, the solubility limit of In in SnTe is relatively high. Typically quoted as $\sim$ 20\%,\cite{Bushmarina1984} our work indicates that impurity phases are present as low as 6.5\%. In either case, these values are considerably higher than that of Tl in PbTe ($\sim$ 1.5\%). In this sense, In-doped SnTe provides access to higher dopant concentrations, potentially enabling study of the competing effects associated with either charge order or coherence beyond a simple single ion picture (a regime for which theoretical work provides little guidance). As we have previously commented, SnTe is also valuable because it undergoes a structural transition for low enough carrier densities.\cite{muldwater1973,kobayashi1976} PbTe might be regarded as a failed ferroelectric in comparison, with a diverging dielectric constant, but no phase transition.\cite{nimtz} This distinction enables investigation of the effects of suppression of the structural transition on the superconductivity, the results of which have showed that the two are essentially uncorrelated, and hence that the enhanced pairing interaction in In-doped SnTe does not derive from a manifestation of the soft phonon.\cite{myprb} On the negative side, In-doped SnTe appears to be more inhomogeneous than Tl-doped PbTe, at least when prepared following the techniques that we have used here. Although bulk superconductivity can be achieved with relatively narrow superconducting transitions for $x \geq 6.1\%$, transport measurements for intermediate compositions, close to the critical concentration for the onset of resonant scattering, cannot provide a good guide to the presence of bulk superconductivity. And lastly, the more complex behavior of the energy of the indium impurity levels inferred from previous measurements  \cite{Bushmarina1984,Bushmarina1986} makes a systematic quantitative analysis of low-temperature transport data considerably more difficult than for Tl-doped PbTe. 

\section{Conclusions}

Our measurements establish a common set of properties associated with In impurities in SnTe and Tl impurities in PbTe. In both materials, a low temperature resistivity upturn is observed, the functional form of which is suggestive of Kondo physics, but in the absence of discernable magnetic impurities. In both cases, the onset of this behavior is correlated with pinning of the Fermi energy in either In or Tl impurity levels, inferred from analysis of transport data. For Tl-doped PbTe, the onset of anomalous scattering coincides with the onset of superconductivity. For In-doped SnTe, bulk superconductivity above 0.7 K is only observed for In concentrations somewhat above that for which the increased scattering is observed, but we cannot rule out superconductivity below this temperature. In both cases, the temperature and field-dependence of the resistivity upturns are suggestive of charge Kondo behavior, the significance of which is of course that the superconducting pairing mechanism might be associated with valence fluctuations of the impurities. Although Tl-doped PbTe is perhaps the better (simpler) material to study the effects of resonant scattering and charge Kondo behavior, In-doped SnTe is useful in the sense that it provides another candidate system which can be explored in a rather different physical regime corresponding to considerably higher dopant concentrations. The generic behavior observed in these two materials, irrespective of subtleties described above, establishes a simple design criterion which might prove useful in the search for new superconductors. We also hope that these results stimulate theoretical work extending the single-ion charge Kondo model to intermediate concentration regimes.

\section{Acknowledgments}
The authors thank Robert E Jones for technical assistance in EMPA measurements, M. R. Beasley, S. A. Kivelson and J. Schmalian for helpful discussions, and E. C. Samulon for help with preparation of several figures in the manuscript. This work was supported by the Department of Energy, Office of Basic Energy Sciences under contract DE-AC02-76SF00515.

\end{document}